\documentclass[final,5p,times,twocolumn]{elsarticle}
\usepackage{graphicx} %
\usepackage{bm}       
\usepackage{dcolumn}  
\usepackage{amssymb}
\usepackage{amsmath}

\newcommand{\bk}{\bm{k}}

\begin{document}


\begin{frontmatter}

\title{Numerical study of quantum Hall effect 
  in two-dimensional multi-band system: \\
    single- and multi-layer graphene}

\author[ma]{Masao Arai}
\ead{arai.masao@nims.go.jp}
\address[ma]{ Computational Materials Science Center, National Institute
  for Materials Science, Tsukuba, Ibaraki 305-0044, Japan }

\author[yh]{Yasuhiro Hatsugai} 
\ead{hatsugai@sakura.cc.tsukuba.ac.jp}

\address[yh]{ Department of Physics, University of Tsukuba, Tsukuba,
  Ibaraki, 305-8571, Japan }

\begin{abstract} 
  The Chern numbers which correspond to quantized Hall conductance
  $\sigma_{xy}$ were calculated for single- and bi-layer honeycomb
  lattices.  The quantization of $\sigma_{xy}$ occurs in entire energy
  range. Several large jumps of Chern numbers appear at van-Hove
  singularities of energy bands without magnetic fields.  The plateaux
  of $\sigma_{xy}$ are discussed from semi-classical quantization.
\end{abstract}
\end{frontmatter}


\section{Introduction}

The quantum Hall effect (QHE) is one of the most peculiar phenomena in
two-dimensional electron system. The quantized Hall conductance
$\sigma_{xy}$ has been widely accepted as a topological quantity
\cite{thouless1982,streda1982,kohmoto1985,niu1985,
  aoki1986,halperin1982,hatsugai1993,hatsugai1993b}. When the chemical
potential is located within an energy gap, the $\sigma_{xy}$ is given
as
\begin{equation}
  \sigma_{xy} = -\frac{e^2}{h} c_F,
\end{equation}
where $c_F$ is a Chern number, which is an topological integer defined
for occupied states. The $c_F$ has been computed for simple
tight-binding models and exotic nature of electron states under
uniform magnetic field has been revealed.

Recently, the discovery of anomalous QHE in graphene
\cite{novoselov2005,zhang2005} shed a new light to the quantization of
Hall conductance. The $\sigma_{xy}$ of graphene shows quantized
plateaux at $\sigma_{xy} = (2N+1)e^2/h$ per spin, where $N$ is an
integer. This means that only plateaux with odd number appear and the
plateau corresponding to $\sigma_{xy} = 0$ does not occur. This
anomalous QHE originates from the massless Dirac particles realized
at the $K$ and $K'$ points in the Brillouin zone of graphene.


In our previous study \cite{arai2009}, we calculated Chern numbers for
realistic multi-band tight-binding models, which describe electrons on
graphene. We found that anomalous quantization of Hall conductance
persists up to van-Hove singularities, as shown in the previous
calculations using simplified model \cite{hatsugai2006}.  In addition,
the envelope of quantized Hall conductance can be well explained with
semi-classical expressions. We further found that the Onsager's
quantization rule \cite{onsager1952},
\begin{equation}
  \frac{S_i(\varepsilon)}{\Omega_\text{BZ}} = (n + \gamma) \phi,
  \label{eq:onsager}
\end{equation}
can predict the positions of plateaux when the band dispersion is well
described by single band. Here, $\phi$ is magnetic flux penetrating in
a unit cell and $\Omega_\text{BZ}$ is the area of irreducible
Brillouin zone.  For Dirac particle regions, above equation also
reproduces the positions of Landau levels if we choose Maslov index
$\gamma = 0$. As this expression does not rely on the linear
dispersion of Dirac particles, it can be applicable in wide energy
region. Therefore, this expression may be utilized for analysis of QHE
and other phenomena of Landau quantization, \textit{e. g.}, de
Haas-van Alphen effects.

In this paper, we extend our study to modified version of
tight-binding models related to graphene layers.  We study how the
sequence of Chern numbers may be altered and whether semi-classical
rules can be applicable to the extended models.





\section{Calculation Methods}

We describe electrons on honeycomb lattice by tight-binding models.
The spin degeneracy is ignored for simplicity.  The uniform magnetic
field is introduced as phase factors $\theta_{ij}$ of hoppings
so that they satisfy
%
$  \sum_{\text{closed loop}} \theta_{ij} = 2\pi\text{(flux quanta in the loop)}$.
%

As the original translational symmetry is broken under uniform
magnetic field, We must consider magnetic translational symmetry,
which is characterized by the magnetic flux $\phi = B\Omega/\varphi_0$
where $\Omega$ is the area of a unit cell and $\varphi_0 = hc/e$ is
the quantized magnetic flux. When $\phi$ is a rational number $p/q$,
we recover enlarged magnetic unit cell whose area is $q$ times larger
than that of the original cell.  Then, eigenstates are labeled by a
wave-number $\bk$. Thus, an energy band splits into $q$ sub-bands by
the uniform field.  

Although the Chern number $c_F$ is given by a sum of the Chern numbers
of all filled bands \cite{thouless1982}, evaluation of them separately
causes numerical difficulty since one needs to treat many Chern
numbers for the weak field limit, where most of them cancel with
each other.  Physical Chern number of the filled many body state is
stable if the energy gap above the Fermi energy is stable. It is
independent of a possible level crossing at far below the Fermi
energy. Then the non Abelian formulation of the Chern number by
collecting all of the filled bands into a single multiplet $\Psi$ is
quite useful \cite{hatsugai2004,hatsugai2005,fukui2005}.  Then the
Chern number $c_F$ of the multiplet $\Psi$ is calculated by
discretizing the Brillouin zone into mesh $\{\bk _\ell\}$
as \cite{fukui2005}
\begin{eqnarray*}
c_F(\mu) &=&
\frac {1}{2\pi } \sum _\ell F(\bk_\ell) 
\\
F(\bk) &=&\text{Arg}\, u_x(\bk)u_y(\bk+\Delta \bk_x)
[u_x(\bk+\Delta\bk_y)u_y(\bk)] ^*
\\
u_\mu(\bk)  &=&\det [\Psi(\bk) ] ^\dagger \Psi(\bk+\Delta\bk_\mu)
\\
\Psi(\bk)  &=&(|\psi_1(\bk) \rangle ,\cdots, |\psi_M(\bk)\rangle )
\end{eqnarray*}
where $\Delta \bk_\nu$ is a discretized momentum along $\nu$
direction, $| \psi_m(\bk) \rangle $ is a Bloch state of the band index
$m$ and $m=M$ is the highest occupied energy band below the chemical
potential $\mu$. This expression can be understood as a
two-dimensional analogue of the KSV formula \cite{ksv1993}. 

This formulation was successfully applied to quantum Hall effect on a
disorder system\cite{song2007}, graphene\cite{hatsugai2006}.  When the
field $\phi=p/q$ is sufficiently small, $p$ sub-bands are grouped
together in general. These grouped bands correspond to a Landau band
of nearly free electrons.  Hereafter, we treat the field $\phi=1/{q}$,
which enables us to consider the weak-field limit easily.


\section{Honeycomb lattice with staggered on-site energy}

\begin{figure}
  \centering
  \includegraphics[height=4cm]{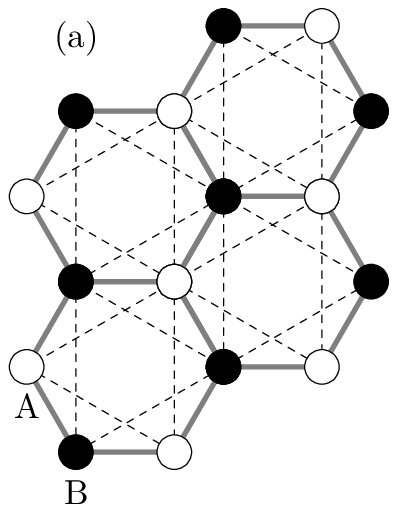}
  \includegraphics[height=4cm]{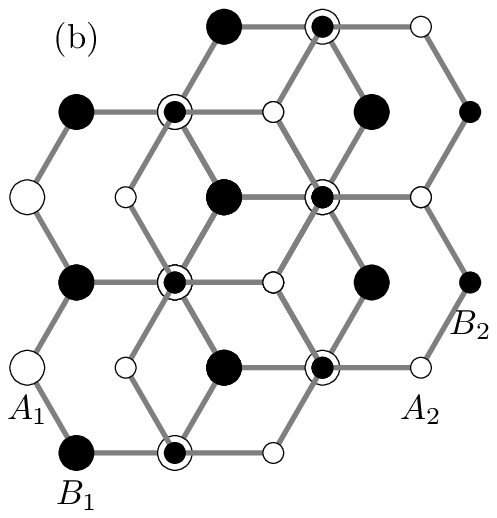}
  \caption{(a) A honeycomb lattice with staggered on-site
    energies. Two inequivalent sites A and B are distinguished by white
    and black circles.  The dashed lines indicate hoppings between
    next-nearest sites.  (b) A structure model of
    bilayer graphene. Carbon atoms on upper and lower layer are
    indicated by small and large circles, respectively.}
  \label{fig:graphene1}
\end{figure}

We treat a honeycomb lattice with two inequivalent sites A and B as
shown in Fig.~\ref{fig:graphene1}(a). Single $\pi$-orbital is put on
each site.  The hoppings between nearest and next-nearest neighbor
sites are included with parameters $t_1$ and $t_2$. The on-site
energies are set to $\pm\Delta$ for A and B sites. Hereafter,we set
$t_1 = 1$, \textit{i.e.}, $t_1$ is chosen as an energy scale.

When $t_2 = \Delta = 0$, this model describes simplified $\pi$-bands
on graphene and the band dispersions which resemble zero-mass Dirac
cones exist at two symmetrically equivalent points, $K$ (1/3 1/3) and
$K'$ (2/3 2/3), in the irreducible Brillouin zone.
The QHE of this model has been extensively studied
theoretically \cite{hatsugai2006}. Here, we briefly explain the
correspondence between quantum mechanical treatment by Chern numbers
and semi-classical quantization rule (\ref{eq:onsager}).

\begin{figure}
  \centering
  \includegraphics[width=8.5cm]{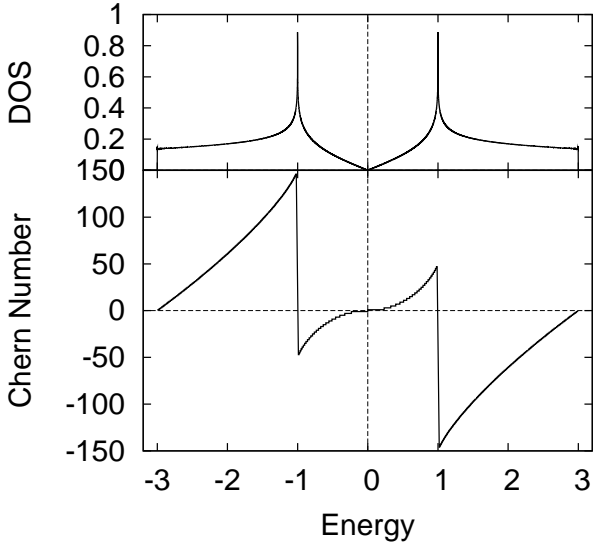}
  \caption{The upper panel shows the total density of states (DOS)
    without magnetic field. The lower panel presents calculated Chern
    number sequence as a function of chemical potential.}
  \label{fig:graphene-qhe}
\end{figure}

The calculated Chern numbers for $\phi = 1/200$ are plotted as a
function of chemical potential in lower panel of
Fig.~\ref{fig:graphene-qhe}. The upper panel shows the total density
of states (DOS) when magnetic field is absent.  From these figures, we
can easily recognize that large discontinuous jumps of Chern numbers
correspond to van-Hove singularities at $\varepsilon = \pm 1$. Within
semi-classical theory, this discontinuity can be explained as a
topological transitions of Fermi surface from electron (hole) pockets
to hole (electron) pockets\cite{arai2009}.

\begin{figure}
  \centering
  \includegraphics[width=8.5cm]{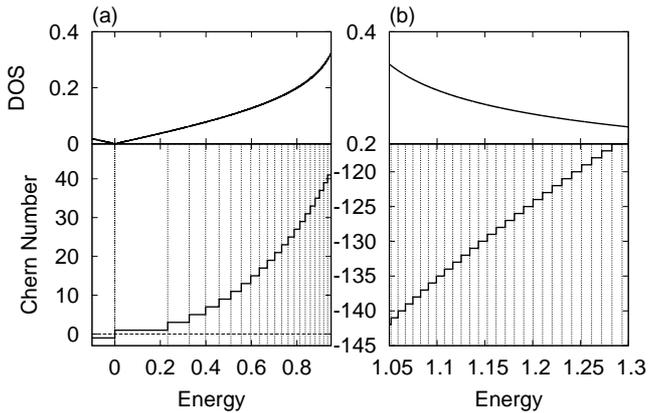}
  \caption{The upper panel shows the total density of states (DOS)
    without magnetic field. The lower panel presents calculated Chern
    number sequence as a function of chemical potential. The vertical
    dashed lines indicate the positions where semi-classical
    quantization rule is satisfied with (a) $\gamma = 0$ and (b)
    $\gamma = 1/2$.}
  \label{fig:graphene-qhe2}
\end{figure}

Fig.~\ref{fig:graphene-qhe2} presents enlarged views of Chern numbers
near the van-Hove singularity at $\varepsilon = 1$.  Anomalous QHE
survives up to van-Hove singularities and the sequence of Chern numbers
is odd number only, while above singularities Chern numbers change
with steps of 1, recovering ordinary QHE \cite{hatsugai2006}. For both
regions, semi-classical quantization rule (\ref{eq:onsager}) can
explain the positions of Landau levels if we choose $\gamma = 1/2$ at
ordinary QHE region and $\gamma = 0$ at anomalous QHE region. This
result is non-trivial near the van-Hove singularity as the band
dispersion around $\varepsilon = 1$ is significantly deformed from the
continuum expressions of massless Dirac $\varepsilon \sim |\bm{k}|$ or
normal effective mass theory $\frac{\bm{k}^2}{2m^{*}}$.

\begin{figure}
  \centering
  \includegraphics[width=8.5cm]{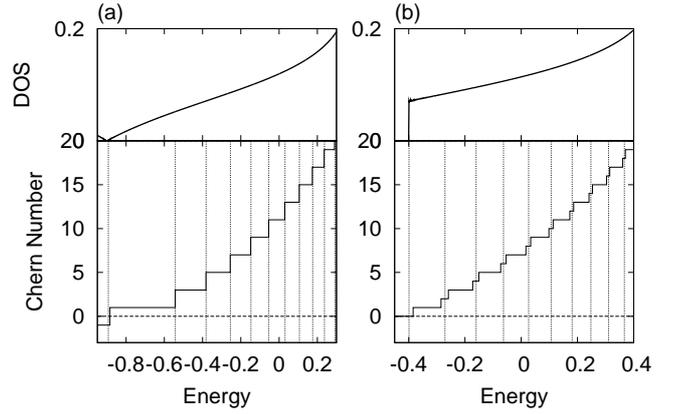}
  \caption{The total DOS and Chern numbers for a honeycomb lattice
    model. The magnetic field is $\phi = 1/101$.  The dashed lines
    indicate the positions where semi-classical quantization rule
    is satisfied with $\gamma = 0$. (a) $\Delta = 0$, $t_2 =
    0.3$. (b) $\Delta = 0.5$, $t_2 = 0.3$.}
  \label{fig:hex2}
\end{figure}

We extend our calculations by incorporating next nearest neighbor
$t_2$ and staggered on-site energy $\pm \Delta$.  When $\Delta = 0$,
the hoppings $t_2$ alone do not destroy the massless Dirac cones.
In this case, the calculated Chern number sequence remains with $c_F =
\cdots, -5, -3, -1, 1, 3, 5, \cdots$ as shown in
Fig.~\ref{fig:hex2}(a).  The positions of Landau levels are still
explained with the semi-classical quantization with $\gamma = 0$.

With finite $\Delta$, the Dirac cones at $K$ and $K'$ acquire band
gaps of order $2\Delta$, which means that Dirac particles have finite
``mass''. When the chemical potential is located above the gap, Fermi
surface is still constructed by two electron pockets around $K$ and
$K'$. In addition, the electronic states around $K$ and $K'$ are
related by lattice symmetry.  If we interpret this band structure
naively, we may expect that two ordinary electron pockets generate
Landau levels at same energies and that Chern numbers between adjacent
energy gaps differ by 2. However, we found that such interpretation
does not hold for current model. As plotted in Fig.~\ref{fig:hex2}(b),
additional plateaux appear in $\sigma_{xy}$ and the Chern numbers
increase with steps of 1. In addition, the positions of Landau levels
do not agree with semi-classical rule(\ref{eq:onsager}) with $\gamma
= 0$ nor $1/2$.  These results mean that electron pockets near $K$
and $K'$ generate Landau levels at different energy positions even
though they are equivalent when magnetic field is absent. Indeed, the
positions of Landau levels may be explained if we choose different
$\gamma$ for Fermi surface segments around K and K'.

\section{Bilayer graphene}



The graphene bilayer consists of two honeycomb lattices as shown in
Fig.~\ref{fig:graphene1}(b).  Each layer is composed of two
inequivalent sites which we denote $A_1$ and $B_1$ on lower layer, and
$A_2$ and $B_2$ on upper one. The $B_2$ sites on upper layer are
located above the $A_1$ sites on lower layer. The largest hoppings
$\gamma_0$ between $\pi$-orbitals are between nearest-neighbor atoms
on the same layer. The inter-layer hoppings are included between $A_1$
and $B_2$ sites with parameter $\gamma_1$. Other hoppings are ignored
for simplicity. These parameters are taken from previous estimations
\cite{charlier1991} as $\gamma_0 = 2.6 eV$ and $\gamma_1/\gamma_0
\approx 0.14$.

The QHE of bilayer graphene was observed experimentally
\cite{novoselov2006} and its electronic structure has been studied
extensively\cite{mccann2006,guinea2006,koshino2006}. Within the
present model, the electron-hole symmetry is maintained and band
dispersions are symmetric around $\varepsilon = 0$. The inter-layer
hoppings modify the Dirac cones which exist in single-graphene and
results in two branches of energy bands for $\varepsilon > 0$. Both
branches behave as $\varepsilon = \delta k^2$ near the $K$ and $K'$
points. In addition, the lower branches remain gapless.

\begin{figure}
  \centering
  \includegraphics[width=8.5cm]{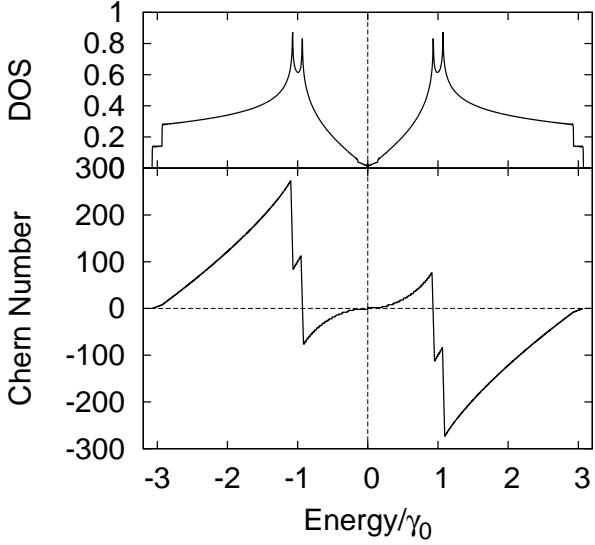}
  \caption{The total DOS (upper panel) and Chern
    number sequences for bilayer graphene are shown.}
  \label{fig:bilayer-qhe}
\end{figure}

\begin{figure}
  \centering
  \includegraphics[width=8.5cm]{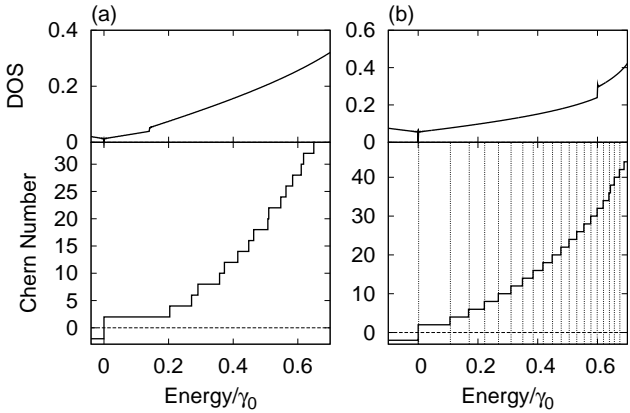}
  \caption{The total DOS and Chern numbers for bilayer honeycomb
    lattice.  Parameters are chosen as (a) $\gamma_1/\gamma_0 = 0.14$
    and (b) $\gamma_1/\gamma_0 = 0.6$. The vertical lines indicate the
    positions of Landau levels calculated from modified semi-classical
    rule (\ref{eq:onsager}).}
  \label{fig:bilayer-qhe2}
\end{figure}

The uniform magnetic field is applied along perpendicular to layers.
Fig.~\ref{fig:bilayer-qhe} shows global structure of calculated Chern
numbers as a function of chemical potential. The total DOS without
magnetic field is also plotted with same energy scale. Because two
branches of energy bands exist, pair of van-Hove singularities appear
around $\varepsilon = \pm 1$. Accordingly, the large discontinuous
jumps of Chern numbers around $\varepsilon = \pm 1$ split into two
jumps. At $\varepsilon = 0$, the Chern numbers jump from -2 to 2,
suggesting the existence of 4 Landau levels at $\varepsilon = 0$. The
positions of other Landau levels shown in Fig.~\ref{fig:bilayer-qhe2}(a)
seem to be located without concrete rule. This is because both upper
and lower branches of energy bands contribute to the quantized Landau
levels. Nonetheless, the Chern numbers always jump by 2, which results
in the sequence of $c_F = \cdots, -4, -2, 2, 4,\cdots$.

We studied Landau levels from lower branch of energy bands by lifting
the upper branch with large $\gamma_1$. Fig.~\ref{fig:bilayer-qhe2}(b)
shows the sequence of Chern numbers with $\gamma_1/\gamma_0 =
0.6$. With this parameter, only lower branch with gapless dispersion
contributes to quantized Landau levels near $\varepsilon = 0$. As a
result, the appearance of Landau levels becomes systematic but can not
be explained by standard semi-classical rule irrespective to the
choice of $\gamma$. In the continuum limit, analytical expression of
Landau levels has been obtained as $\varepsilon \sim \sqrt{n(n-1)}$
\cite{mccann2006}.  Using this expression as a guide, we have tried
several modifications to the semi-classical rule and found that the
equation
\begin{equation}
  \frac{S_i(\varepsilon)}{\Omega_\text{BZ}} = \sqrt{(n(n-1)} \phi
  \label{eq:onsager2}
\end{equation}
agrees with calculated Landau levels within the energy scale presented
in Fig.~\ref{fig:bilayer-qhe2}(b).  However, we have not found any
theoretical support for above equation. Further studies
would be necessary to clarify the quantization rule.

\section{Conclusion}

We studied quantum Hall effect on single and bilayer honeycomb
lattices. For single-layer honeycomb lattice without staggered on-site
energies, the semi-classical theory explains the position of Landau
level if we choose $\gamma$ appropriately. When the on-site energies
are introduced, additional plateaux will appear. This indicate that
valley degeneracy between $K$ and $K'$ is broken when the magnetic
field is applied.

For bilayer graphene, the Chern number sequence of $c_F = \cdots, -4,
-2, 2, 4, 6, \cdots$ is obtained, which agrees with previous
theoretical results. The semi-classical interpretation is difficult
when multiple bands contribute to the same energy regions.


This research by MA is partially supported by a Grant-in-Aid for
Scientific Research, No.~18540331 and No.~17064016 from JSPS.  The
work by YH is also supported in part by Grants-in-Aid for Scientific
Research, No. 20340098 and No. 20654034 from JSPS, No. 220029004
(physics of new quantum phases in super clean materials) and 20046002
(Novel States of Matter Induced by Frustration) on Priority Areas from
MEXT.



\end{document}